\documentclass[prl,aps,amssymb,showpacs,twocolumn]{revtex4}
\usepackage{graphicx}
\begin{document}

\tolerance 10000


\title{Resonant spin polarization in a two-dimensional hole gas:
Effect of the Luttinger term, structural inversion asymmetry and
Zeeman splitting}

\author {Tianxing Ma\footnote{txma@pks.mpg.de}}

\affiliation{Max Planck Institute for the Physics of Complex
Systems, D-01187 Dresden, Germany\\}
\begin{abstract}
\begin{center}

\parbox{17cm}{The electric-field-induced resonant spin polarization of a two-dimensional
hole gas described by Luttinger Hamiltonian with structural
inversion asymmetry and Zeeman splitting in a perpendicular magnetic
field was studied. The spin polarization arising from splitting
between the light and the heavy hole bands shows a resonant peak at
a certain magnetic field. Especially, the competition between the
Luttinger term and the structural inversion asymmetry leads to a
rich resonant peaks structure, and the required magnetic field for
the resonance may be effectively reduced by enlarging the effective
width of the quantum well. Furthermore, the Zeeman splitting tends
to move the resonant spin polarization to a relative high magnetic
field and destroy these rich resonant spin phenomena. Finally, both
the height and the weight of the resonant peak increase as the
temperature decreases. It is believed that such resonant spin
phenomena can be verified in the sample of a two-dimensional hole
gas, and it may provide an efficient way to control spin
polarization by an external electric field.}

\end{center}
\end{abstract}

\maketitle

\section{Introduction}
The physics of spin-orbit coupled system has attracted great
attention in the filed of spintronics. In particular, the spin-orbit
coupling allows for manipulation of the electron spin via electric
field, rather than magnetic field, creating the potential for
applications in areas from spintronics to quantum
computing\cite{Apl,Sarmarth,Rmp}. Theoretically, the spin-orbit
coupling reveals fundamental physics related to topological phases
and their applications to the intrinsic and quantum spin Hall
effect\cite{Sci,Sinova,Kane,Bernevig,Bernevigs,Qi}. Experimentally,
electric-induced spin accumulation has been reported in an electron
doped sample with the use of Kerr rotation
microscopy\cite{Kato1,Kato2} and in a two-dimensional hole gas
(2DHG) by angle-resolved polarization detection\cite{Wunderlich}.

To identify the intrinsic spin Hall effect in experiments, resonant
intrinsic spin Hall conductance has been predicted by several
authors\cite{Shenrc,Dai,Mar}. In a perpendicular magnetic field, the
resonance effect in the two-dimensional electron gas (2DEG) stems
from energy crossing of different Landau levels near the Fermi level
due to the competition of Zeeman energy splitting and Rashba
spin-orbit coupling\cite{Shenrc}, while in the hole-doped system,
the resonant intrinsic spin Hall conductance is due to the
transition between mostly spin-$-\frac{1}{2}$ holes and
spin-$\frac{3}{2}$ holes\cite{Mar}. Even in the absence of a
magnetic field, the Rashba term induces an energy level crossing in
the lowest heavy hole subband, which gives rise to a resonant spin
Hall conductance in a 2DHG\cite{Dai}. However, there have not yet
been experimental reports on the observation of the resonant spin
Hall effect or related phenomena, which is likely due to the
combination of the difficulty in detecting the spin current or spin
accumulation in the high magnetic field and the lack of experimental
efforts in looking into these phenomena\cite{Fczhang}.

Spin polarization induced by electric fields or currents has been
proposed in the spin-orbit coupled systems\cite{Sp1,Sp2,Tao,Shenrp},
and several experiments have been devoted to generate spin
polarization in semiconductors with spin-orbit coupling\cite{spe}.
Very recently, electric-field induced resonant spin polarization was
predicted in a 2DEG\cite{Shenrp}. It was found that a tiny electric
field may generate a finite spin polarization in a disordered Rashba
system in the presence of a magnetic field. As a result, the
electric spin susceptibility exhibits a resonant peak when the Fermi
surface goes through the crossing point of two Landau levels, which
provides a mechanism to control spin polarization efficiently by an
electric field in semiconductors. As the spin polarization can be
measured very accurately, it is believed that the effect can be
verified in the samples of a 2DEG\cite{Shenrp}.

In this paper, we study the resonant electric-field-induced spin
polarization of a 2DHG in detail, which has some intriguing and
observable physical consequences. The general form to describe the
spin transport in a 2DHG is the Luttinger model\cite{Luttinger1956}
with Rashba spin-orbit coupling arising from the structural
inversion asymmetry (SIA)\cite{Mar,Dai,Zhang,SIA}, and such a system
has recently been realized in several experimental
studies\cite{Wunderlich,SIAE}. When a magnetic field is present, the
most general Hamiltonian should involve spin Zeeman terms. However,
the Land$\acute{e}$ g factor may reduce its absolute value, pass
through \emph{zero} or even change sign under a hydrostatic
pressure\cite{Zero1,Zero2,Zero3}, and electrical transport
measurements under hydrostatic pressure have been performed in the
limit of vanishing Land$\acute{e}$ g factor in previous
experiments\cite{Zero1,Zero2,Zero3}. In the presence of a
perpendicular magnetic field, we find that the spin polarization
arising from splitting between the light and the heavy hole bands
shows a resonant peak at a certain magnetic field. Especially, the
competition between the Luttinger term and the Rashba spin-orbit
coupling leads to a rich resonant peaks structure, and the required
magnetic field for the resonance may be effectively reduced by
enlarging the effective width of the quantum well. However, the
Zeeman splitting tends to move such resonant spin polarization to a
relative high magnetic field and destroy these rich resonant spin
phenomena. Finally, both the height and the weight of the resonant
peak increase as the temperature decreases, and the effect of
disorder is discussed. As the spin polarization can be measured very
accurately it is believed that this effect can be verified in the
sample of a 2DHG, and it may provide an efficient way to control
spin polarization by an external electric
field\cite{Shenrp,Fczhang}.

\section{Theoretical framework}
Our starting Hamiltonian for a 2DHG in a magnetic field $B\hat{z}$
is a sum of Luttinger, spin-$\vec{S}$=$\frac{3}{2}$ SIA and the
Zeeman terms\cite{Zhang,Luttinger1956,SIA,Dai,Mar}:
\begin{eqnarray}
H&=&\frac{1}{2m} (\gamma_1 + \frac{5}{2} \gamma_2) \Pi^2 -
2\frac{\gamma_2}{m} (\Pi\cdot S)^2 +\alpha (\vec{S} \times \Pi)
\cdot{\hat{z}} \nonumber \\
&-&\kappa\frac{e\hbar}{mc}S\cdot B
\end{eqnarray}where $\Pi$=$P-\frac{e}{c}A$ is the mechanical momentum, $e$=$-|e|$ is the electric
charge for an electron, $m$ is the bare electron mass, and $\alpha$
is the Rashba spin-orbit coupling. In addition, $\gamma_{1}$ and
$\gamma_{2}$ are two dimensionless parameters modeling the effective
mass and spin-orbit coupling around the $\Gamma$ point, and $\kappa$
is the effective $g$-factor. The confinement of the well in the $z$
direction quantizes the momentum along this axis, which is
approximated by the relation $\langle p_z \rangle$=0, ${\langle
p_z^2 \rangle} \approx (\pi \hbar/ d)^2$ for a quantum well with
thickness $d$\cite{Zhang}.

We use the explicit matrix notation with $S$=$\frac{3}{2}$
eigenstates in the order $S_{z}$=$+\frac{3}{2}$, $+\frac{1}{2}$,
$-\frac{1}{2}$, $-\frac{3}{2}$. By introducing the destruction
operator\cite{Luttinger1956}
$a$=$\frac{1}{\sqrt{2m\hbar\omega}}(\Pi_{x}+i\Pi_{y})$, and the
creation operator
$a^{\dag}$=$\frac{1}{\sqrt{2m\hbar\omega}}(\Pi_{x}-i\Pi_{y})$ to
describe the Landau levels, Hamiltonian (1) can be rewritten as
\begin{eqnarray}
&H&=\hbar\omega\left(
\begin{array}{cccc}
H_{11}& i\sqrt{3}\lambda a^{\dag} & -\sqrt{3}\gamma _{2}a^{+2} & 0 \\
-i\sqrt{3}\lambda a & H_{22} & 2i\lambda a^{\dag} & -\sqrt{3}\gamma _{2}a^{\dag2} \\
-\sqrt{3}\gamma _{2}a^{2} & -2i\lambda a & H_{33} & i\sqrt{3}\lambda a^{\dag} \\
0 & -\sqrt{3}\gamma _{2}a^{2} & -i\sqrt{3}\lambda a & H_{44} )
\end{array}
\right) \nonumber \\
&H&_{NN}=[\gamma_{1}-(-1)^{N}\gamma_{2}](a^{\dag }a+\frac{1}{2})+
\frac{\beta}{2}[\gamma_1+(-1)^{N}2\gamma_2]\nonumber \\
&-&(\frac{5}{2}-N)\kappa
\end{eqnarray}
where $N$=1, 2, 3, 4, the dimensionless parameters $\lambda$=$\alpha
m\sqrt{\frac{c}{2\hbar eB}}$,
$\beta$=$\frac{\pi^{2}\hbar}{d^{2}m\omega}$ and the magnetic length
$l_{b}$=$\sqrt{\frac{\hbar c}{eB}}$. The corresponding eigenvectors
are expressed as
\begin{equation}
\left\vert n,s,f\right\rangle =\left(
\begin{array}{c}
C_{nsf1}\phi _{n} \\
C_{nsf2}\phi _{n-1} \\
C_{nsf3}\phi _{n-2} \\
C_{nsf4}\phi _{n-3}
\end{array}%
\right),
\end{equation}%
where $\phi _{n}$ is the eigenstate of the n$th$ Landul level in the
absence of the spin-orbit coupling, and $n$ is a non-negative
integer. In a large $n$ limit, we can deduce that states $\left\vert
n,+1,\pm1\right\rangle$ indicate light-hole bands and $\left\vert
n,-1,\pm1\right\rangle$ indicate heavy-hole bands\cite{Zhang,Mar}.
We should add that when $n<3$, the definition of $\left\vert
n,s,f\right\rangle $ is not exact, so we simply take $\left\vert
2,-1,1\right\rangle$ as the lowest energy level of $n$=2 and
$\left\vert 1,1,-1\right\rangle$ indicates the lowest energy level
of $n$=1 in the whole paper.

\begin{figure}
\includegraphics[scale=0.575]{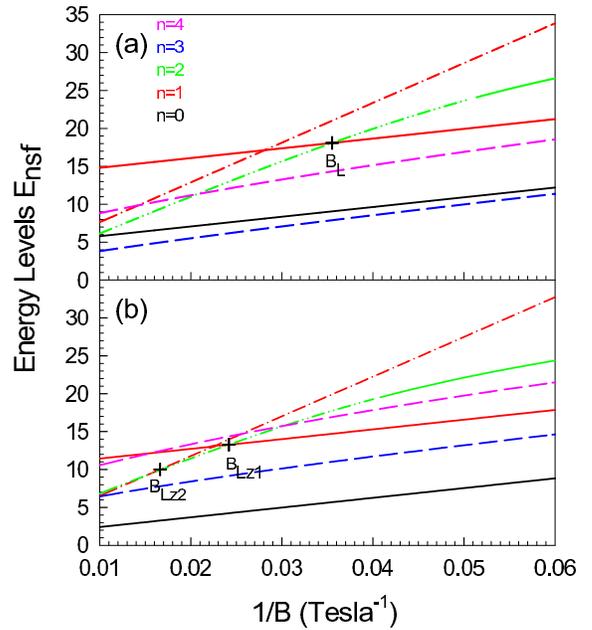}
\caption{(Color online) Landau levels (units: $\hbar\omega$) as
functions of $1/B$. Different colors denote different n and only
energy levels no higher than the energy level of resonant point are
shown. Solid lines indicate Landau levels of mostly
spin-$\frac{3}{2}$ holes, dashed lines indicate Landau levels of
mostly spin-$-\frac{3}{2}$ holes, dash-dot lines indicate Landau
levels of mostly spin-$\frac{1}{2}$ holes, dash-dot-dot lines
indicate Landau levels of mostly spin-$-\frac{1}{2}$ holes. (a)
$\kappa$=0; (b) $\kappa$=2.25. }
\end{figure}

If we apply a weak in-plane electric field in the $y$-direction,
then the electric spin susceptibility can be evaluated by the Kubo
formula in the linear response theory\cite{Mahan}
\begin{eqnarray}
X^{\alpha
y}_{E}&=&\frac{e\hbar}{L_{x}L_{y}}\mathbf{Im}\sum_{nsf,n^{\prime
}s^{\prime }f^{\prime }}\frac{(f_{n^{\prime }s^{\prime }f^{\prime
}}-f_{nsf})}{\epsilon _{nsf}-\epsilon _{n^{\prime }s^{\prime
}f^{\prime }}} \nonumber \\
&\times &\frac{ \left\langle n,s,f\right\vert S_{\alpha}\left\vert
n^{\prime },s^{\prime },f^{\prime }\right\rangle\left\langle
n^{\prime },s^{\prime },f^{\prime }\right\vert
\upsilon_{y}\left\vert n,s,f\right\rangle }{\epsilon _{nsf}-\epsilon
_{n^{\prime }s^{\prime }f^{\prime }}+i/\tau}, \nonumber \\
f_{nsf}&=&\frac{1}{e^{(\epsilon _{nsf}-\mu)/k_{B}T}+1},
\end{eqnarray}
where $\mu$ is the chemical potential, $\epsilon_{nsf}$=$\hbar\omega
E_{nsf}$ is the eigenvalues within Eq. (1), and $v_{y}$ is the
velocity in $y$-direction. From the Kubo formula (4), we can see
that only $n'$=$n\pm 1$ contributes to the spin susceptibility. In
particular, it is natural to point out that if $\epsilon
_{nsf}$=$\epsilon _{n\pm 1s'f'}$ happens near the Fermi energy and
for a long lifetime $\tau$, a divergent $X^{\alpha y}_{E}$ may
appear. Resonant spin phenomenon means an intriguing and observable
physical consequence in
experiments\cite{Shenrp,Mar,Dai,Fczhang,Mas}. To be convenient for
future experimental detection, we will discuss the effect of
Luttinger term, SIA term, Zeeman splitting, and temperature on this
resonant spin phenomenon in detail.

\section{Energy level depending on the Luttinger term, structural inversion asymmetry and
Zeeman splitting}

The properties of energy spectrum depending on magnetic field
determine the behavior of spin transport. To further our story step
by step, we study the energy levels as functions of the inverse of
magnetic field within Eq. (1) when $\alpha$=0 firstly. Depending on
the confinement scale $d$ the Luttinger term is dominant for $d$ not
too small, while the SIA term becomes dominant for infinitely thin
wells. Moreover, to learn the effect of Zeeman splitting on this
resonant spin phenomenon, we distinguish Fig.1 (a) with $\kappa$=0
from Fig.1 (b) with $\kappa$=2.25\cite{Luttinger1956,Winkler}. Other
parameters used are the same, $\gamma_{1}$=6.92, $\gamma_{2}$=2.1
and $d$=8.3nm\cite{Wunderlich,Zhang}.

We use lines with different colors to denote different $n$. In order
to give a more clear illumination bellow, we only plot lines within
energy levels no higher than the energy level of the resonant point,
which shall contribute to the spin transport. As we have discussed,
if the energy level crossing between states $\left\vert
n,s,f\right\rangle$ and $\left\vert n\pm1,s',f'\right\rangle$ occurs
near Fermi energy, it may lead to a resonance. Though there are
energy crosses that may lead to resonances when $1/B<0.01$
Tesla$^{-1}$ theoretically, the corresponding magnetic field is
unavailable experimentally. Moreover, there are no energy cross when
$1/B>0.06$ Tesla$^{-1}$ for present parameters, so we only consider
the case when $0.01$ Tesla$^{-1}$$<1/B<0.06$ Tesla$^{-1}$.

In Fig.1 (a), the energy cross between states $\left\vert
1,1,-1\right\rangle$ and $\left\vert 2,-1,1\right\rangle$ occurs at
$B_{L}$=28.25 Tesla ( marked by a cross ). For a set of sample's
parameters, the behavior of energy levels depends on the magnetic
field. Whether this energy cross shall lead to a resonance at the
corresponding magnetic filed, is determined by the hole density,
which means that the ``effective" energy cross shall appear near the
Fermi energy, and this can be related directly to the filling
factor, $\nu $=$\frac{N_{h}}{N_{\phi}}$=$\frac{n_{h}2\pi\hbar
c}{eB}$. As shown in Fig.1 (a), the realization of resonance
requires that $3<\frac{n_{h}2\pi\hbar c}{eB_{L}}<4$, and the hole
density shall be $2.07\times 10^{16}$/m$^{2}$$< n_{h}<2.75\times
10^{16}$/m$^{2}$. Including the effect of Zeeman splitting, as shown
in Fig.1 (b), the effective energy cross moves to a relative higher
magnetic field, $B_{Lz1}$=41.46 Tesla ( marked as a cross ), and the
required hole density for the resonance shall be
$2<\frac{n_{h}2\pi\hbar c}{eB_{Lz1}}<3$, $i.e$, $2.01\times
10^{16}$/m$^{2}$$< n_{h}<3.01\times 10^{16}$/m$^{2}$.

The energy cross occurring at $B_{Lz1}$ means that
$E_{1,1,-1}$=$E_{2,-1,1}$. When $\alpha$=0, an analytical equation
can be derived from $E_{1,1,-1}$=$E_{2,-1,1}$, which is
\begin{eqnarray}
B_{Lz1}d^{2}=\frac{\pi ^{2}\hbar c}{e}\frac{%
4\gamma _{1}\gamma _{2}+4\gamma _{2}^{2}}{\gamma _{1}^{2}+3\gamma
_{1}\gamma _{2}+8\gamma _{2}^{2}-2\kappa (\gamma _{1}+\gamma _{2})}
\label{BrLz1},
\end{eqnarray}
where we know that $B_{Lz1}$ increases as $\kappa$ increases.
However, the Zeeman splitting introduce another resonant point at
$B_{Lz2}$, which is due to the energy cross between sates
$\left\vert 1,1,1\right\rangle$ and $\left\vert
2,-1,1\right\rangle$, namely, $E_{1,1,1}$=$E_{2,-1,1}$. The required
magnetic field
\begin{eqnarray}
B_{Lz2}d^{2}=\frac{\pi ^{2}\hbar c}{e}\frac{4\gamma _{2}}{2\gamma
_{1}+3\gamma _{2}-\kappa -\frac{6\gamma _{2}^{2}}{\kappa
}}\geq0.853\frac{\pi ^{2}\hbar c}{e}\label{BrLz2},
\end{eqnarray}
and the equal sign is satisfied when $\kappa$=$\sqrt{6}\gamma _{2}$.
The resonance at this point is introduced by the zeeman splitting
since $\kappa$=0 is excluded from this equation. Moreover, $B_{Lz2}$
is determined by the competition between the Luttinger term and the
Zeeman splitting, and $B_{Lz2}$ decreases as $\kappa$ increases when
$\kappa<\sqrt{6}\gamma _{2}$. From Eq. (5) and (6), it is useful to
find that the required magnetic field for the resonance may be
effectively reduced by enlarging the effective width of the quantum
well.

\begin{figure}
\includegraphics[scale=0.575]{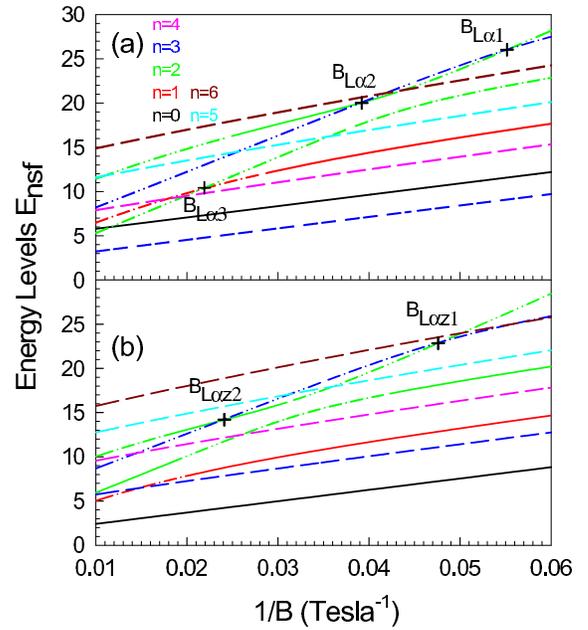}
\caption{(Color online) Caption is the same as that in Fig.1 except
(a) $\alpha$=10$^5$ m/s, and $\kappa$=0; (b) $\alpha$=10$^5$ m/s,
and $\kappa$=2.25. }
\end{figure}
Secondly, let us discuss the effect of SIA on this resonant spin
phenomenon. The relatively large 5 meV measured
spitting\cite{Wunderlich,Zhang} of the heavy hole band implies that
the effect of Rashba spin-orbit coupling arising from the SIA term
is important. Energy levels as functions as $1/B$ with
$\alpha$=$10^{5}$m/s\cite{Wunderlich,Zhang} are shown in Fig.2, and
we distinct (a) from (b) in case of $\kappa$=$0$ and
$\kappa$=$2.25$. The key points that may lead to resonant spin
transport have been marked as crosses in Fig. 2.

Comparing energy levels in Fig.1 (a) with those in Fig.2 (a), the
SIA term moves the energy crosses ( at $B_{L\alpha3}$ ) between
states $\left\vert 1,1,-1\right\rangle$ and $\left\vert
2,-1,1\right\rangle$ to a relative high magnetic field. However, a
new set of energy crosses ( at $B_{L\alpha1}$ and $B_{L\alpha2}$ )
appear in relative low magnetic fields, which are due to states
$\left\vert 2,1,-1\right\rangle$ and $\left\vert
3,-1,1\right\rangle$, and there are at least three energy crosses
which may lead to resonant spin transport. The first resonant point
appears at $B_{L\alpha 1}$=$18.09$ Tesla, which requires that
$7<\frac{n^{1}_{h}2\pi\hbar c}{eB_{L\alpha 1}}<8$. The second
resonant point appears at $B_{L\alpha 2}$=$25.70$ Tesla, requiring
that $6<\frac{n^{2}_{h}2\pi\hbar c}{eB_{L\alpha 2}}<7$. Since the
properties of energy levels are depending on $B$ through
$\lambda$=$\alpha m\sqrt{\frac{c}{2\hbar eB}}$ and
$\beta$=$\frac{\pi^{2}\hbar c}{d^{2}eB}$, the new set of resonant
points can be related to the competition between the SIA and the
Luttinger term. It is interesting to point out that, if the range of
$n^{1}_{h}$ and $n^{2}_{h}$ has some conjunct values, a rich
resonant peaks structure of spin transport shall appear, and we will
discuss this intriguing case bellow.

The energy levels as functions of $1/B$ when $\kappa$=2.25 have been
shown in Fig. 2 (b). There are two effective energy crosses.
However, the first resonant point appears at $B_{L\alpha z2}$=20.85
Tesla, which requires that $7<\frac{n^{z1}_{h}2\pi\hbar
c}{eB_{L\alpha z1}}<8$, and the second resonant point appears at
$B_{L\alpha z2}$=42.18 Tesla, requiring $7<\frac{n^{z2}_{h}2\pi\hbar
c}{eB_{L\alpha z2}}<8$. The Zeeman splitting tends to move the
resonant points to a higher magnetic field and resonance in a sample
with a higher hole density is required.

\begin{figure}
\includegraphics[scale=0.575]{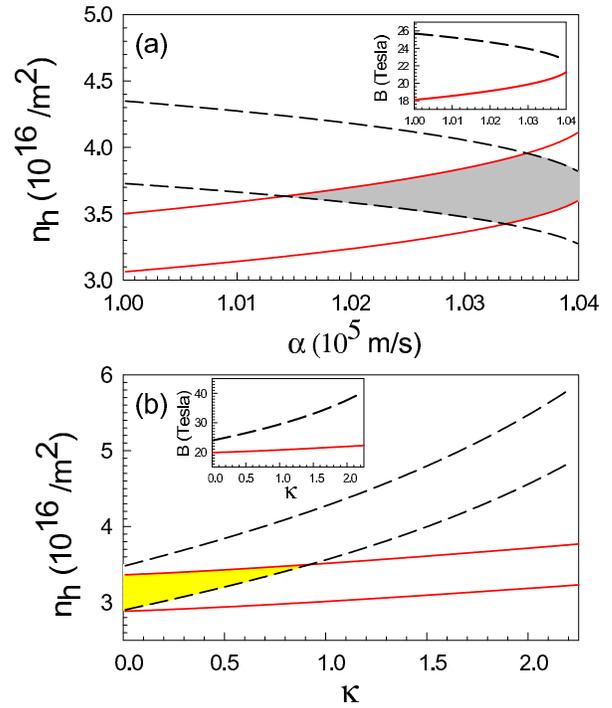}
\caption{(Color online) The required range of hole density for
resonance at the relative high ( area between dark dash lines ) and
low ( area between red lines ) magnetic field as functions of (a)
$\alpha$ when $\kappa$=0; (b) $\kappa$ at
$\alpha$=$1.03\times10^5$/m$^2$. Inset: magnetic fields at two
different resonant points as function of (a) $\alpha$ when
$\kappa$=0; (b) $\kappa$ at $\alpha$=$1.03\times10^5$/m$^2$. }
\end{figure}

Thirdly, to learn an overall understanding on the resonance
depending of SIA term, we plot the required range of hole density
for resonance as functions of $\alpha$ in Fig. 3 (a), and the
magnetic field at resonant point as functions as $\alpha$ has been
shown in the inset. These resonances are due to energy crosses of
states $\left\vert 2,1,-1\right\rangle$ and $\left\vert
3,-1,1\right\rangle$, namely, $E_{2,1,-1}$=$E_{3,-1,1}$. The
required range of hole density for resonance at the relative high
magnetic field ( range between dash dark lines ) decreases as
$\alpha$ increases. However, the required range of hole density for
resonance at the relative low magnetic field ( range between red
lines ) increases as $\kappa$ increases, and there is a conjunct
scope ( gray area in Fig. 3 (a) ), which shall lead to a rich
resonant peaks structure for a sample by changing the magnetic
field.

Let us take a look on energy levels in Fig.1 (a) and Fig.1 (b)
together, as well as those in Fig.2 (a) and Fig.2 (b). It seems that
larger effective $g$-factor $\kappa$ will move such resonant spin
transport to a higher magnetic field. To learn more on this aspect,
we study the effect of $\kappa$ on the required hole density for
resonance, as well as magnetic field at resonant point. These
resonances are due to energy crosses of states $\left\vert
2,1,-1\right\rangle$ and $\left\vert 3,-1,1\right\rangle$, and
parameters used are $\gamma_{1}$=6.92, $\gamma_{2}$=2.1, $d$=8.3nm,
and $\alpha$=$1.03\times 10^5$ m/s. As shown in Fig. 3 (b), the
required range of hole density for resonance at the relative high
magnetic field ( range between dash dark lines ) increases more
quickly than the required range of hole density at the relative low
magnetic field ( range between red lines ) as $\kappa$ increases,
which removes the conjunct scope ( yellow area in Fig. 3 (b) ), and
the rich resonant spin transport of a sample will disappear as
$\kappa$ is large enough.

\section{Resonant spin susceptibility}
\begin{figure}
\includegraphics[scale=0.575]{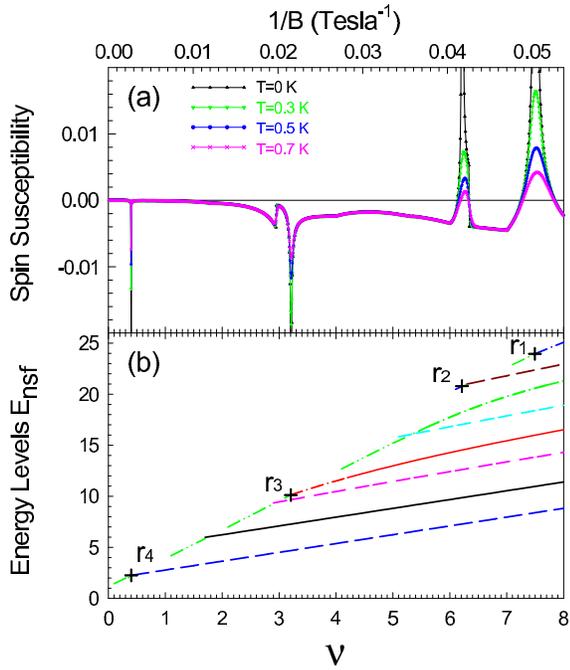}
\caption{(Color online) (a) Resonant spin susceptibility versus
$1/B$ (or $\nu$) at several temperature for weak electric fields.
Parameters used are the same as those in Fig.2 (a) except
$\alpha$=$1.03\times 10^{5}$m/s. (b) Landau levels as functions of
filling factors $\nu$. Different colors denote different n and only
energy levels occupied for corresponding $\nu$ are shown.}
\end{figure}

Now let us turn to study the resonant spin susceptibility. Our
numerical result for $X^{yy}_{E}$ has been shown in Fig.4 (a), and a
remarkable rich resonant peaks structure appears, which indicate
that a weak field may induce an intriguing and observable physical
consequence of a 2DHG in the presence of a perpendicular magnetic
field. Since the value of $\kappa$ can be reduced by using
hydrostatic pressure\cite{Zero1,Zero2,Zero3}, we take $\kappa$=0
without loss of generality. Other parameters used are
$n_{h}$=$3.6\times 10^{16}$/m$^{2}$, $\gamma_{1}$=6.92,
$\gamma_{2}$=2.1, $d$=8.3nm, and $\alpha$=$1.03 \times 10^5$m/s. The
magnetic fields for resonances are respectively $B_{r1}$=19.87
Tesla, $B_{r2}$=23.96 Tesla, and $B_{r3}$=46.42 Tesla, which are in
the range of present experimental capability, and $B_{r4}$ is rather
high so it need not be considered. Every energy cross for the
resonance has been marked as $r1$, $r2$, $r3$ and $r4$ in Fig.4 (b),
and a careful analysis reveals that the resonance at $r1$ and $r3$
are contributed from the transition between mostly
spin-$-\frac{1}{2}$ and mostly spin-$\frac{1}{2}$ holes, and the
resonance at $r2$ is due to the interplay between mostly
spin-$-\frac{1}{2}$ and mostly spin-$\frac{3}{2}$ holes. As the spin
polarization can be measured very accurately, it is believed that
this effect can be verified in the samples of a 2DHG. Temperature is
another important factor on this resonant spin polarization. In Fig.
4 (a), we have also plotted resonant spin susceptibility at several
temperatures. As we can see, both the height and the weight of the
resonant peak increase as the temperature decreases at low
temperature.

\begin{figure}
\includegraphics[scale=0.64]{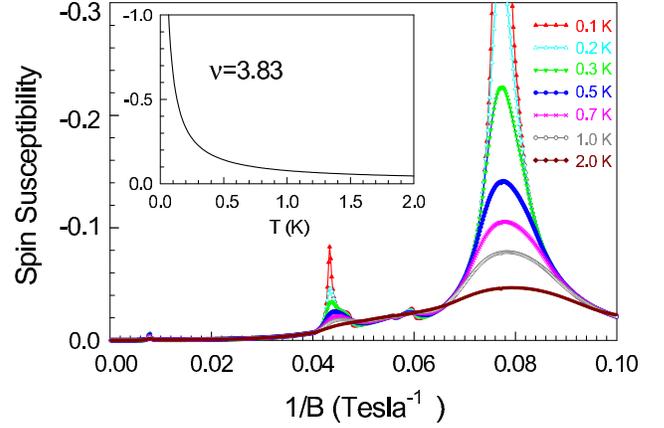}
\caption{(Color online) Resonant spin susceptibility ( units:
$\hbar/4\pi l^{2}_{b} N/C$ ) versus $1/B$ at several temperature.
The parameters are the same as those in Fig. 4 except $d=13$ nm and
$n_{h}$=$1.2\times 10^{16}$/m$^{2}$. In the inset, temperature
dependence of the height of the resonance peak is plotted. }
\end{figure}

As we have discussed, the required magnetic field for the resonance
may be effectively reduced by enlarging the effective width of the
quantum well. To be convenient for future experimental detection,
and learn more on the effect of temperature, we show resonant spin
susceptibility at several temperatures for a relative low magnetic
field in Fig. 5. The resonance appears at about 12.96 Tesla and the
peak is still prominent even at 0.5 K. In the inset of Fig. 5, we
show the temperature dependence of the height of the resonant peak.
The characteristic temperature for the occurrence of the peak can be
estimated to be about 2 K at the resonant field for the parameters
in the caption.

We have assumed no potential disorder in our theory. The effect of
disorder in the 2DHG with spin-orbit coupling, especially in a
strong magnetic field, is not well understood at this
point\cite{Dis1,Dis2}. However, the effect of disorder on such kind
of resonant spin phenomena in a 2DEG has been discussed by Bao and
Shen\cite{Shenrp} most recently. Their numerical simulation
illustrated that impurity potential opens an energy gap near the
resonant point and suppressed the effect gradually with increasing
strength of disorder. Although the resonant spin phenomena in a 2DHG
is much richer and more complicated, the essential nature of
resonance is the same as the case in a 2DEG, which is caused by the
energy crossing between different Landau levels. Moreover, in the
absence of a magnetic field, numerical study of the spin transport
in the Luttinger model indicates that the spin transport in the weak
disorder regime remain almost the same as the value for the pure
system\cite{Dis2}. It seems to be reasonable to assume that resonant
spin polarization in a 2DHG shall survive in the weak disorder
regime.

\section{Summary}
In summary, we have studied the electric-field-induced resonant spin
polarization of a 2DHG within the Luttinger model with structural
inversion asymmetry and Zeeman splitting in a perpendicular magnetic
field. The spin polarization arising from splitting between the
light and the heavy hole bands shows a resonant peak at a certain
magnetic field, and a rich resonant peaks structure is predicted,
which is due to the competition between the Luttinger term and the
structural inversion asymmetry. The required magnetic field for the
resonance may be effectively reduced by enlarging the effective
width of the quantum well. However, the Zeeman splitting tends to
move the resonant spin polarization to a relative high magnetic
field and destroy this rich resonant peaks structure. Finally, the
resonant value of the electric spin susceptibility decay with the
temperature. Our calculations show that the parameters ( the
magnetic field, the effective $g$-factor, the hole density, the well
thickness, and the Rashba spin-orbit coupling strength ) for the
resonance are likely accessible in experiments. It is believed that
such resonant spin phenomena can be verified in the samples of
two-dimensional hole gas, and it provides an efficient way to
control spin polarization by an external electric field.

\acknowledgments We thank Shun-Qing Shen for careful reading and
many helpful discussions. We hank Yun-Juan Bao and Qin Liu for many
helpful discussions.

\clearpage

\end{document}